# Traffic Congestion Awareness and On-Demand Distribution in Vehicular Delay-Tolerant Networks in California I-210 Freeway

Xiaofei Liu
*School of Computer Science*
*University of Nottingham*
Nottingham, United Kingdom
psxxl24@nottingham.ac.uk

Milena Radenkovic
*School of Computer Science*
*University of Nottingham*
Nottingham, United Kingdom
milena.radenkovic@nottingham.ac.uk

***Abstract*—In vehicular networks under edge computing environments, vehicle-to-vehicle delay-tolerant networking (V-DTN) can disseminate congestion warnings to other vehicles via a store-carry-forward mechanism, helping them proactively choose suitable routes. However, most existing in-vehicle information dissemination methods rely on flooding or limited flooding strategies, broadcasting alerts across the entire network whenever congestion is detected. This leads to excessive redundant copies and consumes node cache space.**

**To address this issue, this paper proposes a congestion-event-aware on-demand message dissemination mechanism. By considering the recurrence and duration of congestion, the mechanism suppresses broadcasts of short-lived, self-dissipating congestion events. Based on real-world PeMS data from California's I-210 corridor, we construct an I-210 Freeway Traffic Congestion Use Case. Experiments show that predicting congestion recurrence heavily relies on historical free-flow speed data of lanes. Without incorporating additional feature dimensions, the accuracy of modeling and comparing lane free-flow speeds across different day types outweighs the choice of machine learning models, and such patterns are difficult to reproduce in simulators. Meanwhile, congestion duration prediction serves effectively as the basis for dynamic-threshold on-demand dissemination in V-DTN, which demonstrate that machine learning-based congestion awareness combined with dynamic-threshold on-demand dissemination significantly reduces network resource consumption and network-layer congestion, enhances delivery reliability, and lowers delivery latency.**

## I. Introduction

With the advancement of vehicle-to-everything (V2X) technology, vehicular delay-tolerant networks (V-DTNs) have become a crucial means for disseminating traffic information in road segments with insufficient infrastructure coverage. In highway scenarios, when congestion occurs, V-DTN can propagate warning messages to vehicles that have not yet entered the congested area, enabling them to reroute or prepare in advance, thereby mitigating the spread of congestion. However, most existing vehicular information dissemination methods rely on flooding or limited flooding strategies, broadcasting warnings across the entire network as soon as congestion is detected. Although flooding-based protocols offer high delivery opportunities, simple implementation, and strong robustness, they generate numerous redundant copies, consuming node cache space, wireless bandwidth, and battery energy. More advanced congestion-aware opportunistic routing protocols (such as Cafe [1] and CafRep [2] [3]) primarily focus on network-level performance metrics such as delivery probability, social paths, delay, and overhead. In contrast, these routing protocols pay relatively little attention to the congestion events themselves. Broadcasting alerts for short-term, periodic congestion events can create excessive redundant messages in bandwidth-constrained vehicular networks, leading to buffer overflow, message loss, and self-induced network congestion, ultimately undermining the reliable delivery of critical warnings. This study aims to explore an underutilized opportunity: leveraging predictions of congestion duration and periodicity as thresholds for on-demand message dissemination in V-DTN, enabling selective warning broadcasts.

## II. Related Work

### A. Traffic Congestion Awareness Based on Machine Learning

Machine learning has become essential for predicting traffic congestion duration. In their review, Liu and Shin found that traditional machine learning efficiently processes nonlinear traffic data and scales better to small datasets than computationally demanding deep learning models [4]. For anomaly detection, Kim et al. proposed SMURP, a hybrid framework that reduces forecast errors by supplementing historical machine learning data with traffic simulation during abnormal conditions [5]. Taher et al. further highlighted that while deep and reinforcement learning excel in accuracy, edge computing improves real-time performance and reduces latency by shifting data processing to the front end [6]. Regarding specific incident duration prediction, Hamad et al. demonstrated the high accuracy of classifiers like SVM, KNN, and Gaussian Process models for real-time management [7], [8], while Xie et al. successfully applied CART regression to multi-source sensor data to predict post-incident intervals [9].

Despite machine learning's proven efficiency and the benefits of integrating simulation and edge computing, signifi-cant research gaps remain. Existing studies primarily address non-routine, accident-induced congestion rather than common congestion with multiple coupled causes. Furthermore, many models rely heavily on offline historical data analysis, which cannot accommodate dynamic urban traffic needs. Future research must establish online prediction frameworks combining edge perception and lightweight machine learning to achieve accurate, real-time control of urban traffic congestion.

### B. Gradient Boosting and LightGBM Algorithm

Gradient Boosting is a successful ensemble machine-learning framework for regression and classification [10], [11] that sequentially trains base models correlated with the negative gradients of the loss function [12]. Its fast variant, LightGBM [13], utilizes Gradient-based One-Side Sampling (GOSS) and Exclusive Feature Bundling (EFB) to significantly enhance training speed and model efficiency. While outperforming XGBoost on large datasets, LightGBM is prone to overfitting on smaller ones [14]. This algorithm has been effectively applied in complex scenarios; for instance, hybridizing LightGBM with physical models resolved time-consuming real-time urban flood predictions [15]. Similarly, Berlotti et al. [16] utilized tree-based models, including LightGBM and XGBoost, to successfully predict traffic flow on unmonitored road segments by mirroring patterns from sensor-equipped neighbors. These applications demonstrate that while gradient boosting methods offer exceptional speed and accuracy in industrial, hydrological, and traffic scenarios, pattern anomalies (e.g., holiday traffic) and insufficient datasets can degrade predictive performance, necessitating additional contextual features for improved generalizability.

### C. Non-Recurrent Traffic Congestion

Early researchers primarily modeled congestion by emphasizing its periodicity and using statistical models to decompose total delay into recurrent (normal demand) and non-recurrent (event-related) delay using field measurements [17]. Dowling et al. further categorized recurrent congestion as demand-driven and non-recurrent as caused by incidents like accidents or weather, developing a framework to evaluate these factors and assess traffic system efficiency [18]. As research evolved, spatial-temporal correlations became vital. Li et al. proposed a four-dimensional tensor model (Coupled SBRTF) using PeMS data to detect non-recurrent congestion, outperforming conventional methods by leveraging multivariate spatiotemporal structures [19]. Similarly, Sun et al. introduced the DxNAT deep learning model, which maps traffic conditions to 2D grid pixels to capture spatial network information often lost in traditional segment-level vectors [20].

While methods have advanced from statistical decomposition to high-dimensional tensor modeling and deep learning, limitations remain. Current research primarily focuses on post-event detection and relies heavily on high-quality, labeled data, struggling in complex urban environments with overlapping incidents. Future research must shift from state detection to the proactive forecasting of dynamic evolution trends, spatial diffusion, and duration of non-periodic congestion.

### D. Vehicular Delay-Tolerant Network

Machine learning (ML) integration can significantly improve Delay Tolerant Networks (DTN) [21]. Bhavani et al. [22] showed that ML-based intelligent algorithms enhance DTN data forwarding—particularly in relay node selection and congestion control—over classic protocols when evaluated via simulators like The ONE [23] and NS-3 [24]. Similarly, regarding VANETs [25], Nazib and Moh [26] found that reinforcement learning routing improves adaptability and QoS in dynamic environments. In V-DTNs, Liu et al. [27] utilized ML to automatically select the most suitable routing protocol based on historical features, significantly improving scenario adaptability. While ML successfully utilizes data-oriented techniques to enhance network intelligence and key performance indexes, current ML-based routing generally focuses on network states (e.g., packet loss or buffer overflow) for passive congestion control, ignoring physical events like traffic congestion or natural calamities that dramatically change topology. Relying solely on network-layer metrics hinders adaptation to new opportunistic node distributions and misses valuable context. Therefore, this research captures traffic congestion event prediction in addition to event-level restricted DTN forwarding and flooding, optimizing limited network resources and bandwidth processing suited for DTN scenarios.

### E. Simulation of Urban MObility (SUMO) and The ONE Simulator

Simulation of Urban Mobility (SUMO) [28] is a microscopic open-source traffic simulator used to convert real-world traffic into digital form. Monga and Mehta combined SUMO with OpenStreetMap to build reproducible vehicle mobility models for VANET routing protocol evaluations [29]. Additionally, Haddouch et al. authenticated SUMO's applicability in modeling urban traffic flows and effectively locating congestion points [30]. Thus, SUMO's detailed microscopic trajectory data provides an extremely realistic mobility basis for communication protocol evaluation, while simultaneously supporting accurate traffic flow modeling and supply optimization.

The ONE simulator [23] evaluates protocols in ”store-carry-forward” delay-tolerant networks where traditional end-to-end routing is unviable. In disaster applications, such as the 2015 Nepal earthquake scenario [31], simulations revealed that unlimited replication protocols like Epidemic caused cache overflow (delivery ¡20%), whereas Spray-and-Wait controlled replicas to maintain a 94.6% delivery rate under constrained resources. Similarly, during a 2016 Italy earthquake stadium evacuation [32], limited-replica protocols outperformed unlimited flooding in achieving higher delivery rates, lower delays, and more stable overhead. Furthermore, integrating ONE with the SUMO traffic simulator improves vehicular mobility modeling. Oda et al. [33] utilized SUMO vehicular trajectories within ONE to evaluate VDTN protocols, finding that Epidemic routing compromised reliability and latency was proportional to path length, highlighting the need to evaluate diverse routing protocols and traffic models.

## III. I-210 Freeway Traffic Congestion Use Case Construction

### A. Data Source: Performance Measurement System (PeMS)

Figure 1 shows Performance Measurement System(PeMS), which is a freeway performance measurement system for all of California [34] [35]. It processes real-time data from loop detectors of various traffic management centers (TMC) in California every day. Its system can handle these data in real time, obtain accurate speed estimations through adaptive algorithms, and aggregate lane data into average values at 5-minute intervals [35]. Based on these data, PeMS calculates key traffic performance indicators, including vehicle mileage, vehicle travel time, speed changes, delay time, and travel time. Users can access and use its application through a web browser and download the data. It also includes a series of special road events such as accident data from California freeway Patrol (CHP) [35], lane maintenance data, and weather data, which help users explore the impacts behind various events.

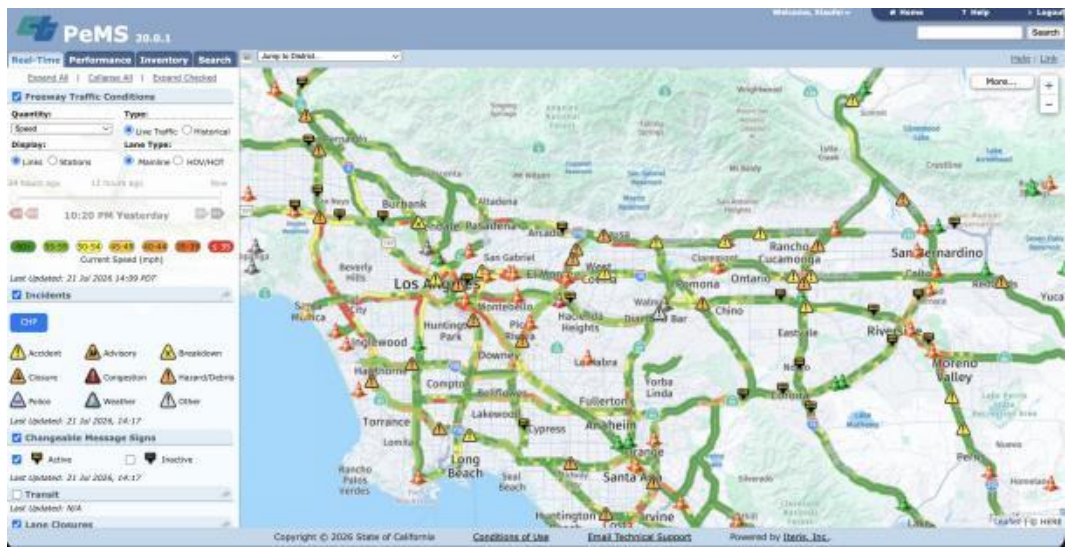

Fig. 1. California Freeway Performance Measurement System

### B. I-210 Freeway Traffic Congestion Dataset Generative Pipeline

This project utilizes real freeway traffic data from the California Department of Transportation's PeMS system, focusing on the bidirectional I-210 Freeway segment in the San Gabriel Valley, Los Angeles County. The study period covers a continuous three-month autumn season from September to November 2024. This timeframe features fully archived and stable-quality data, making it ideal for offline batch processing. Autumn is a typical commuting season with well-defined patterns of recurrent traffic peaks and congestion, while avoiding the impact of summer vacation travel and year-end holidays such as Christmas and Thanksgiving on traffic behavior. Additionally, weather conditions and extreme weather events significantly affect freeway speeds and congestion levels—each additional millimeter per hour of precipitation increases the likelihood of driver deceleration by 5.8% [36]. California experienced no rainfall between September and November 2024 [37], effectively minimizing traffic disruptions caused by rain and reducing the influence of variable weather factors on traffic stability. Furthermore, the three-month duration provides sufficient non-recurrent congestion samples to support the training and evaluation of machine learning models, aligning with recent practices that rely on long-term traffic monitoring data for event and congestion modeling [4]. The study focused on the I-210 Freeway in the 7th zone of California's PeMS (Absolute Postmile: 22.30–44.99).As shown in figure 2, the 5-minute granularity traffic status data from the inner loop detectors in the PeMS database was selected to detect congestion events and extract early features. The sensors of PeMS provided a large amount of data for all monitoring stations in both directions, including station type, coverage length, sample quantity, and observed value percentage, etc. During the data processing, this project only retained the detection stations of the main line (ML) within the corridor range, and conducted quality screening based on the average observed rate (% Observed) of each station over the entire study period. Eventually, 49 valid stations were retained to ensure the reliability and representativeness of the data.

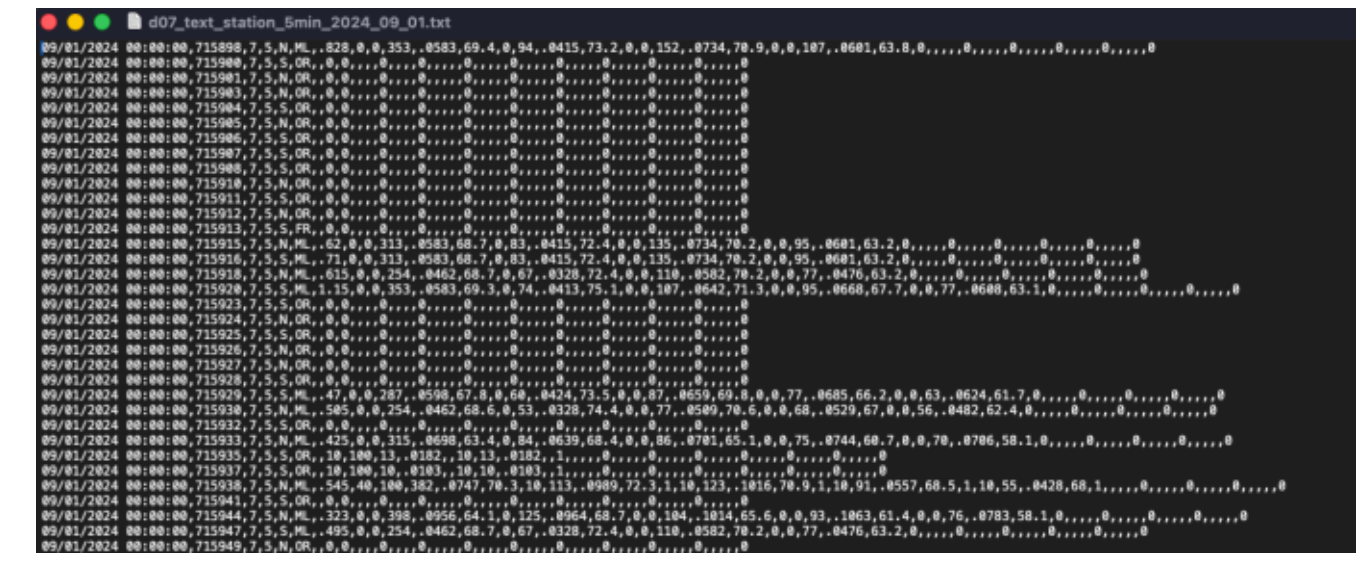

Fig. 2. Original 5-minutes data in I-210 Freeway Traffic Congestion Use Case

To distinguish between recurrent and non-recurrent congestion, this study grouped historical vehicle speeds by station, whether it was a working day, and time period, and took the median and the lower bound of the normal speed band (15th percentile) of the group as indicators to represent the normal speed and its lower limit of normal fluctuation, respectively, to detect the recurrent congestion pattern. Figure 3 shows the calculated median and 15th percentile for a station, if the average vehicle speed of the event was below the lower limit of the same period, it was determined as non-recurrent congestion, otherwise, it was recurrent congestion. Meanwhile, given that congestion is a phenomenon that propagates along roads, the duration of a congestion event also depends on the speed changes upstream and downstream of the congested area. Therefore, during the dataset generation process, to help the model better learn the spatial changes in congestion, average vehicle speed detection data for upstream and downstream stations of the congested station were added. This data comes from the average vehicle speeds of adjacent stations upstream and downstream of the identified congested station within the same time window. The vehicle speed of downstream neighboring stations helps in bottleneck identification; worsening congestion in these areas prevents timely vehicle evacuation, leading to increased congestion time. The vehicle speed of upstream neighboring stations helps in identifying whether there is a continuous increase in traffic flow and worsening congestion; a decrease in upstream vehicle speed indicates a tendency for congestion to spread downstream. Subsequently, the congestion events were detected. When the vehicle speed was lower than 35 mph, it was determined as the start of congestion, and when it rose to 40 mph, it was determined as the end [38]. Events with a duration of less than 15 minutes were excluded. Figure 4 shows the flowchart of dataset generation. Based on this method, a total of 5693 congestion events were identified. Finally, the features within a 10-minute window after the event start were extracted, with the congestion duration as the regression target and whether it was non-recurrent congestion as the classification target, to construct the final training data set. The features and labels of the final data set are shown in Table I.

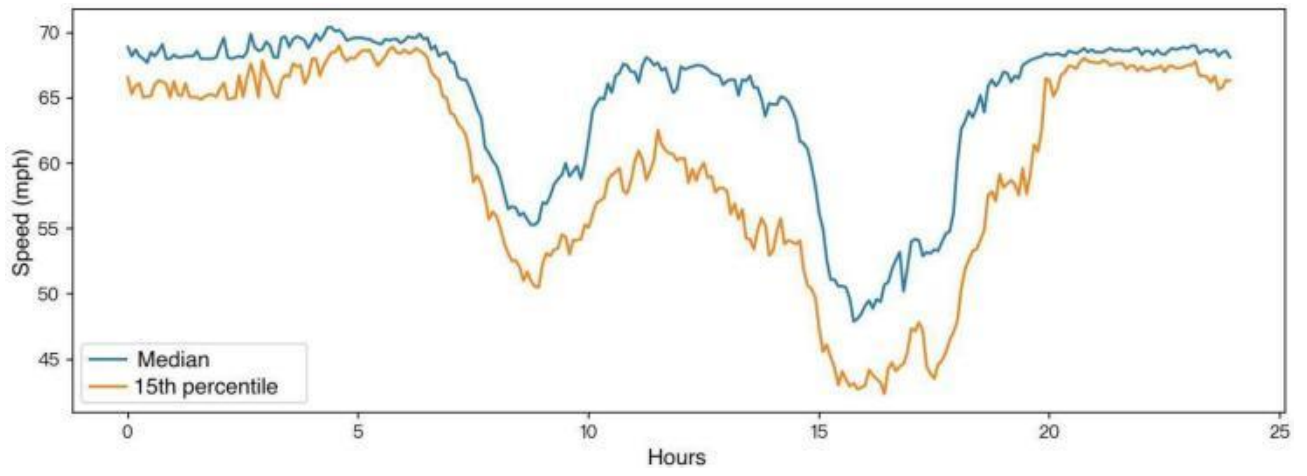

Fig. 3. Detect Station NO.770578 Workday Baseline in I-210 Freeway Traffic Congestion Use Case

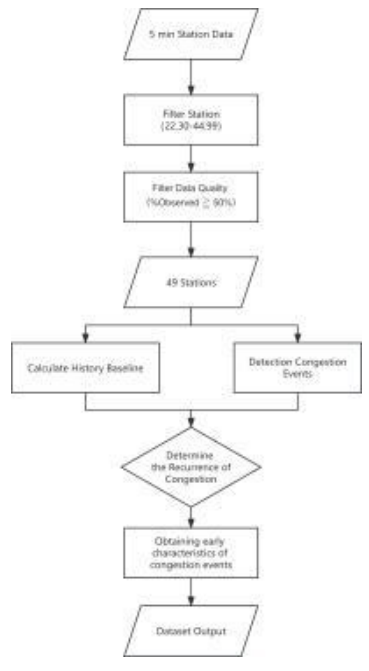

Fig. 4. Dataset generation pipeline in I-210 Freeway Traffic Congestion Use Case

| Use Case Features | Explaination |
|---|---|
| station | PeMS Station ID |
| speed_onset | The car speed at the moment of event beginning (mph) |
| speed_min | Minimum speed within the early event window(mph) |
| speed_mean | Average speed within the early event window(mph) |
| speed_drop_rate | Vehicle speed decrease rate (mph/min) |
| flow_onset | The traffic flow at the moment of event beginning |
| flow_min | Early event average flow |
| hour | The congestion event occurred time of the day |
| up_speed_mean | Upstream event window average speed |
| dn_speed_mean | Downstream event window average speed |
| base_low | 15th percentile speed |
| duration_min | Event total duration (min) |
| is_nonrecurrent | Recurrent congestion determine |

TABLE I
FEATURES IN I-210 FREEWAY TRAFFIC CONGESTION USE CASE DATASET

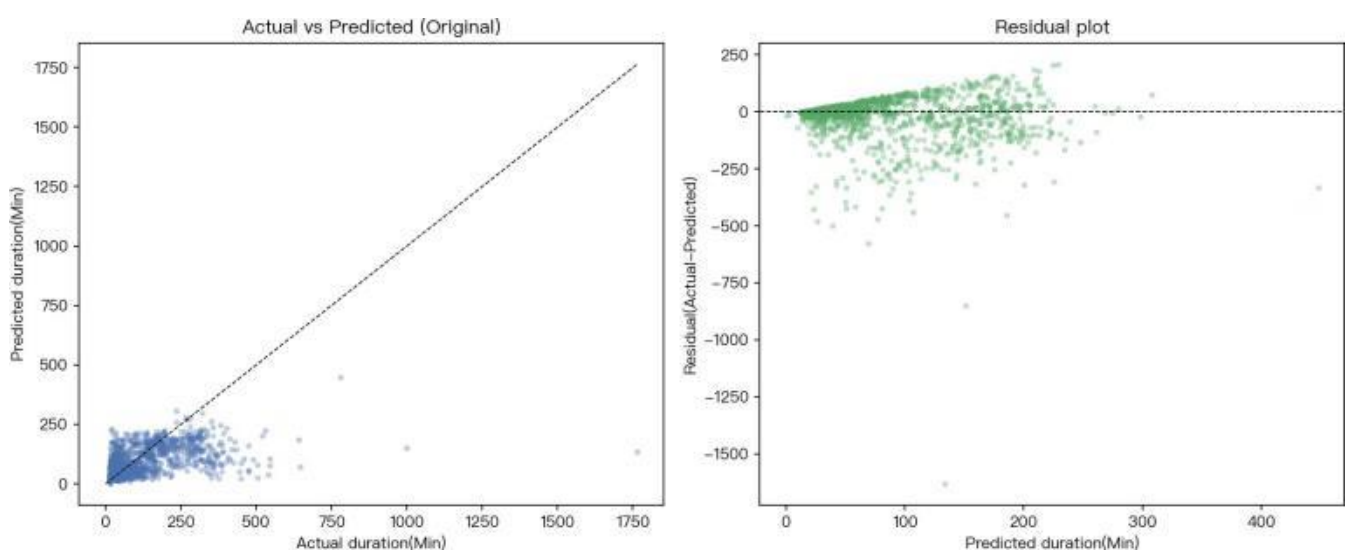

Fig. 5. Residual analysis chart on Congestion Duration I-210 Freeway Traffic Congestion Use Case Dataset (Original data)

## IV. MACHINE LEARING MODEL TRAINING

### A. LightGBM Regression: Predicting congestion duration

Figure5 shows the residual plot of the unprocessed full-sample training model. In the regression task, the long-tail effect is fully manifested in the model without limiting the duration of congestion events. In the fullsample model training, the MAE reaches 64.3 and the RMSE reaches 113.1. The piecewise error analysis in Figure 6 shows that the MSE for events of 0–60 minutes is 32.1, while the MSE for extremely long events exceeding 240 minutes reaches 203.1. These extreme events account for only about 3% but contribute the vast majority of the total error; therefore, the modeling range is limited to events ≤ 240 minutes, covering approximately 97% of the total data.

Based on the above data analysis, a new regression analysis was performed on the optimized dataset. The results and residual analysis plots are shown in Figure 7 and Table II. The regression task achieved 5-Fold cross-validation within the limited range with a MAE of 34.7 ± 1.1 and an RMSE of 52.6, significantly better than the full-sample data. Figure 8 shows the feature importance analysis, the incorporation of upstream and downstream spatiotemporal features ranked first and second in feature importance. This result proves that the prediction of congestion duration is inherently more important at the spatial level. Downstream station speeds reflect the existence of bottlenecks ahead and determine whether traffic flow can be dispersed, while upstream station speeds reflect whether there is continuous inflow from behind and determine whether queuing will worsen. Early observations from individual stations, together with spatiotemporal information from adjacent stations, contribute to the model's good predictive ability.

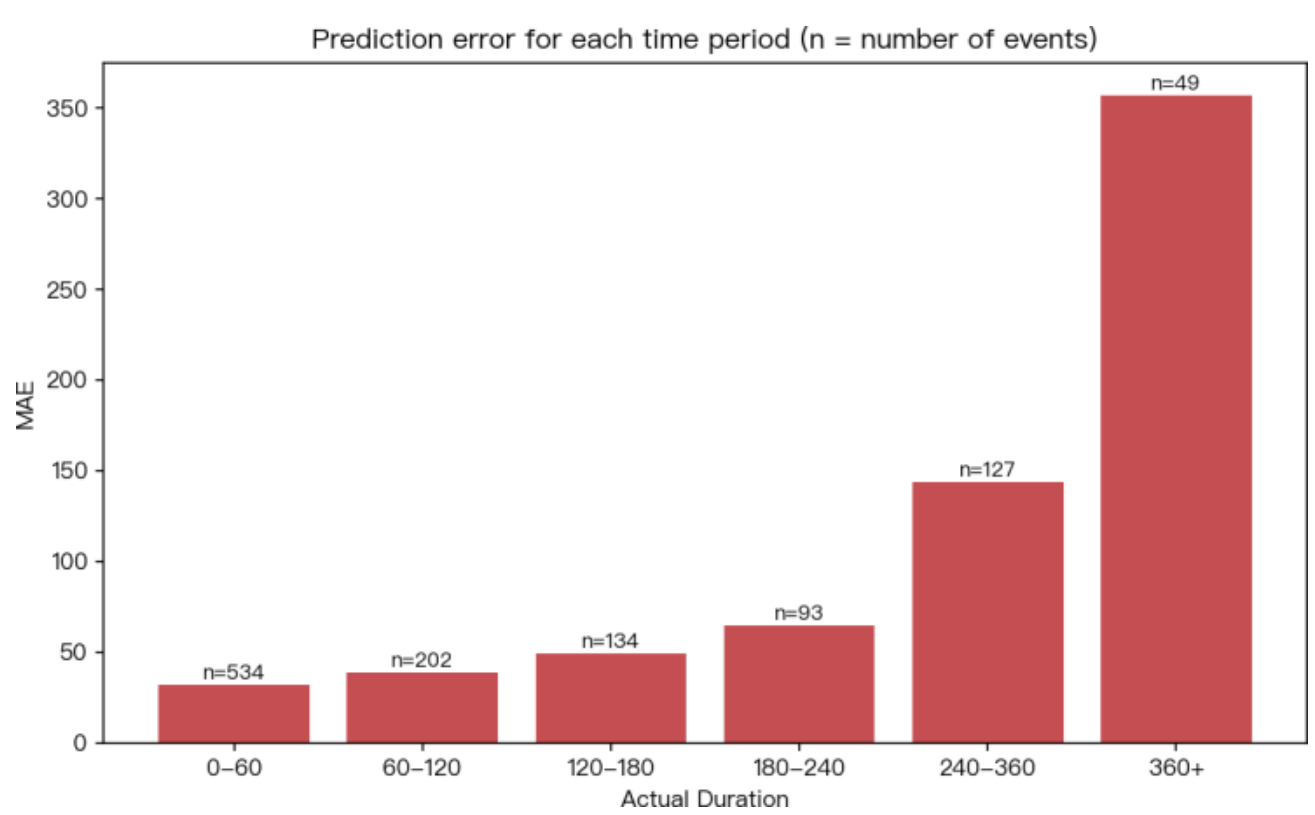

Fig. 6. Correlation between prediction error and actual duration On I-210 Freeway Traffic Congestion Use Case Dataset

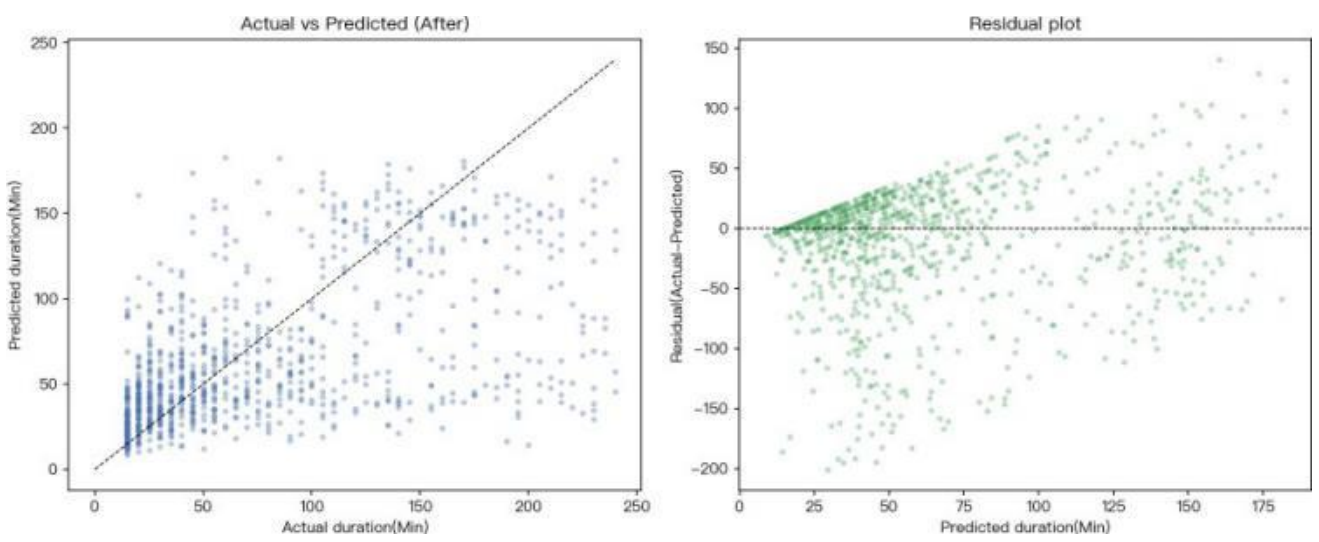

Fig. 7. Residual analysis chart on Congestion Duration of I-210 Freeway Traffic Congestion Use Case Dataset (Duration ¡ 240)

| Evaluation parameters | Original Data | Duration < 240 | Duration < 180 |
|---|---|---|---|
| RMSE | 113.1 | 52.6 | 38.1 |
| 5-Fold average MAE | 65.4 ± 2.9 | 34.7 ± 1.1 | 26.7 ± 0.7 |

TABLE II
PERFORMANCE COMPARISON BETWEEN THE ORIGINAL MODEL AND THE LONG-TAIL-REMOVED MODEL

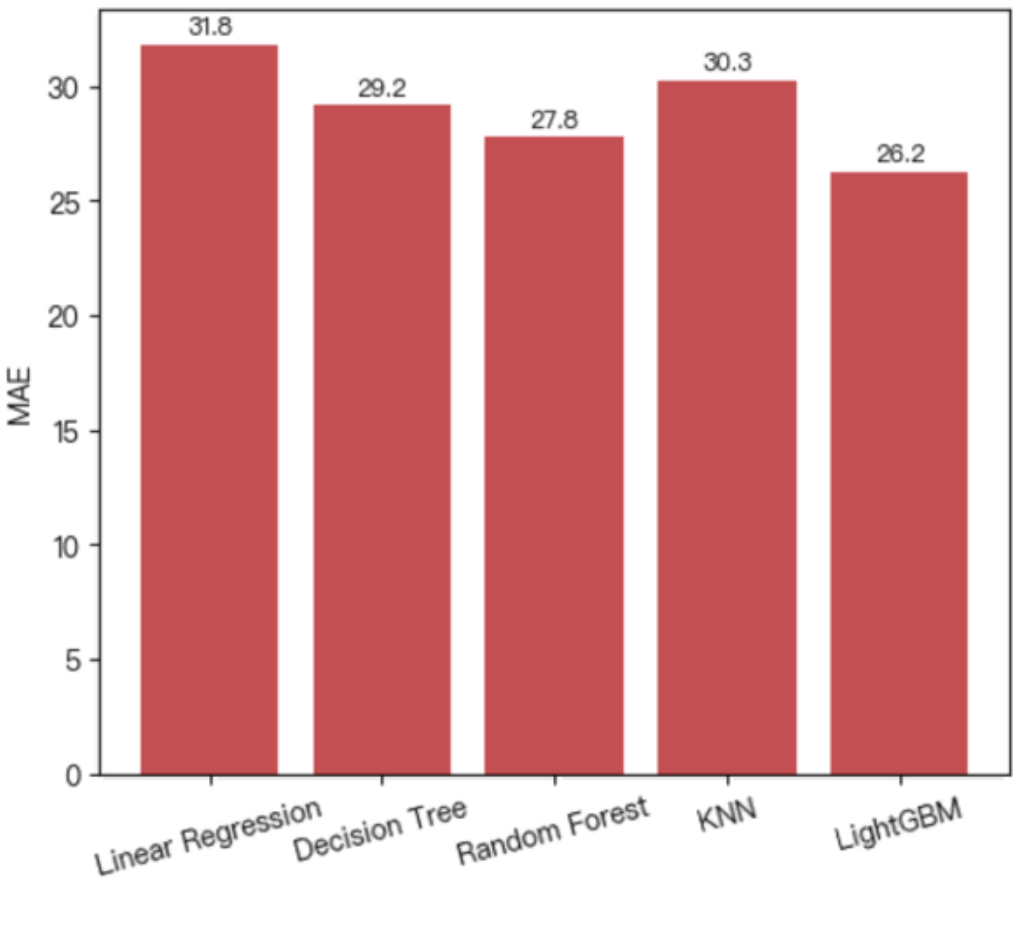

Fig. 9. Comparison of Regression Algorithm Performance in I-210 Freeway Traffic Congestion Use Case Dataset

Figure 9 illustrates the performance of different models in predicting congestion duration within the I-210 Freeway Traffic Congestion Use Case. LightGBM achieved the lowest MAE, consistent with our predictions in Section 4.1. Random Forest ranked second, highlighting the advantage of ensemble learning in this use case—ensemble methods typically outperform individual base models in complex regression tasks due to their superior nonlinear fitting capability and robustness. As an efficient gradient boosting model, LightGBM delivered the best results in this task, indicating that the dataset involves intricate feature interactions. LightGBM's histogram-based algorithm and leaf-wise growth strategy effectively capture these complex patterns. Linear Regression had the highest MAE and performed the worst, suggesting significant nonlinear relationships between features and the target variable, leading to substantial underfitting for models assuming strict linearity—a characteristic consistent with traffic congestion dynamics. KNN performed only slightly better than linear regression, likely due to the "curse of dimensionality" associated with high-dimensional data. Overall, tree-based ensemble methods significantly outperformed traditional linear and distance-based algorithms on this dataset and regression task. Among the five candidate models, LightGBM emerged as the optimal choice.

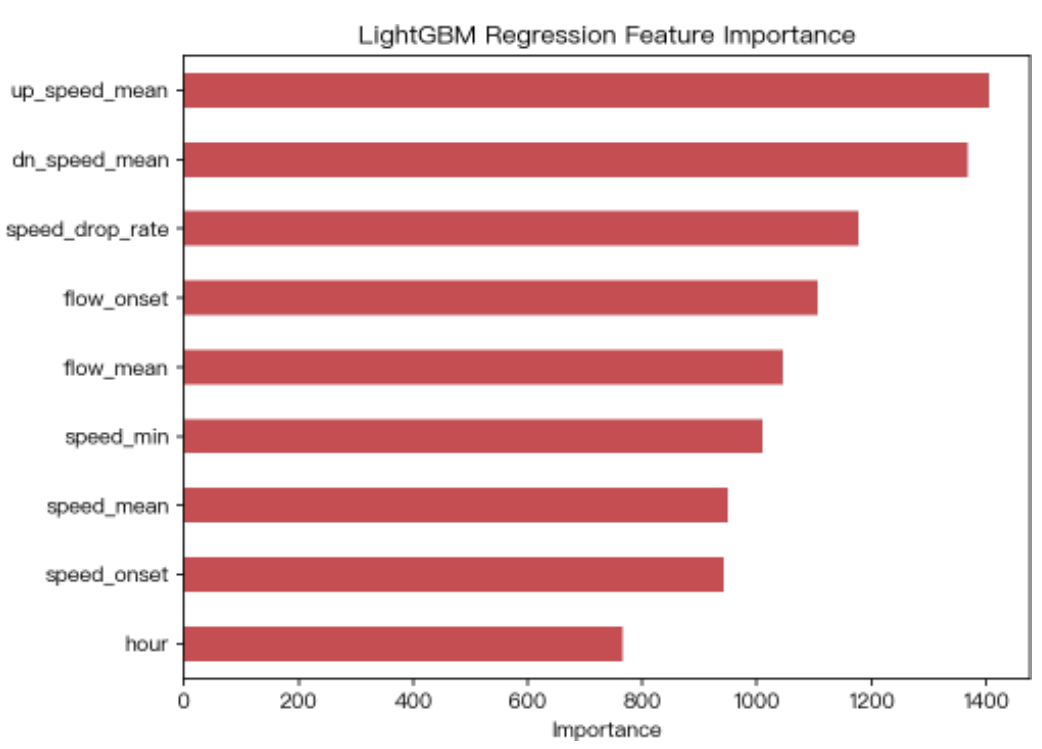

Fig. 8. I-210 Freeway Traffic Congestion Use Case Dataset Regression Feature Importance

### B. Classification: Non-Recurrent Congestion Determine

To distinguish whether a single congestion event belongs to non-recurrent congestion, this study first attempts to construct multiple machine learning classification models for discrimination, with experimental results shown in Figure 10. The results indicate that the overall performance differences among various models are not significant, suggesting that under the current feature system and data scale, increasing model complexity does not lead to noticeable improvements in classification accuracy. Therefore, a feature importance analysis (based on LightGBM classification) was conducted on the dataset, as shown in Figure 11. Further analysis revealed that the variable "lower bound of the 15th percentile of free-flow speed" plays a dominant role in classification outcomes. All models primarily rely on this speed threshold feature during discrimination, rather than fully capturing more complex spatiotemporal traffic evolution patterns.

Based on the above findings, this study directly adopted a rule-based comparison method, that is, directly comparing the free-flow vehicle speed with the vehicle speed when a congestion event occurs. The results are shown in Table III. Results demonstrate that this method achieves performance comparable to machine learning models. This suggests that, given the current data conditions, the advantages of machine learning models are not fully realized; classification performance mainly stems from the free-flow speed features themselves, rather than from the models' deep understanding of recurrent and non-recurrent congestion mechanisms. Clearly, identifying and predicting recurring congestion requires high standards for the completeness and temporal coverage of historical data. Reliable detection of recurring congestion typically demands long-term traffic operation data spanning several years, and modeling should be conducted by grouping data according to factors such as weekdays, weekends, holidays, seasonal variations, and special events to extract consistently recurring congestion patterns. The experiments in this study used only data from September to November 2024, a relatively short time span that is insufficient to fully reflect traffic characteristics across different months, seasons, and holidays. Therefore, the assessment and prediction of recurring congestion presented in this paper remain limited. Future research should incorporate historical data over longer time periods and adopt more refined grouping methods to enhance the reliability of recurring congestion identification.

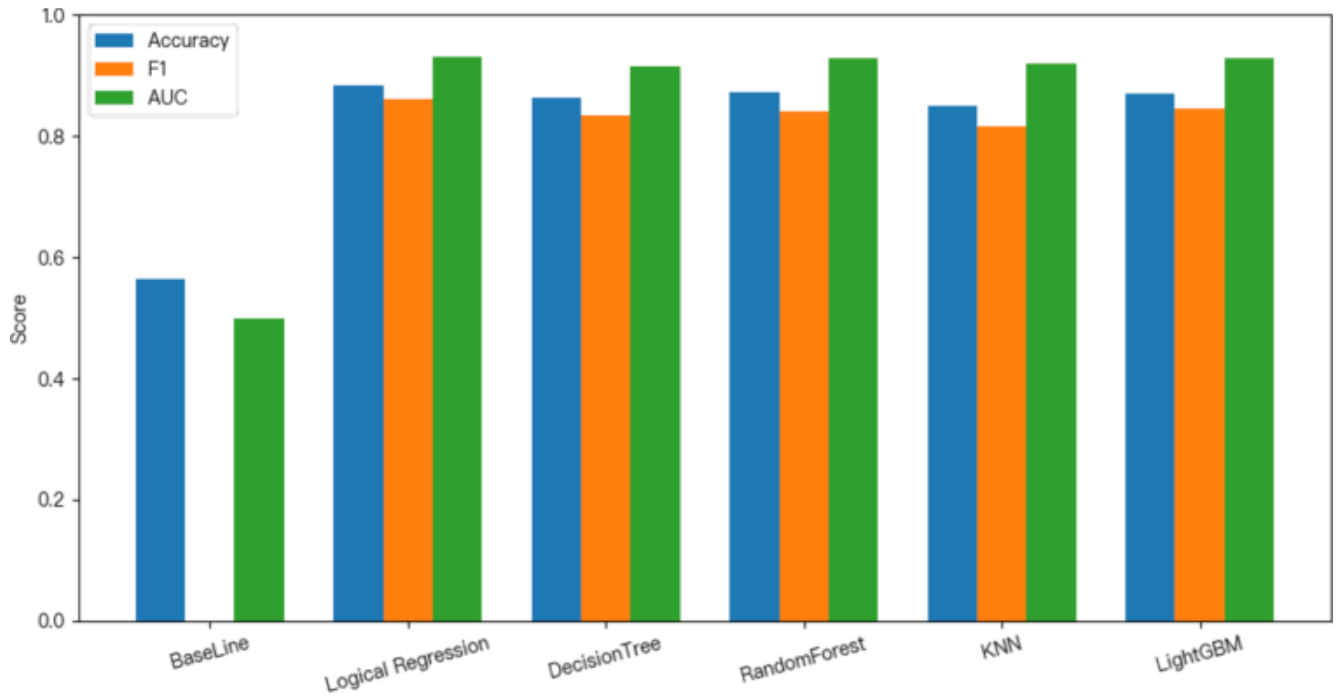

Fig. 10. Accuracy, F1-score and AUC in Non-Recurrent Congestion Classification for I-210 Freeway Traffic Congestion Use Case

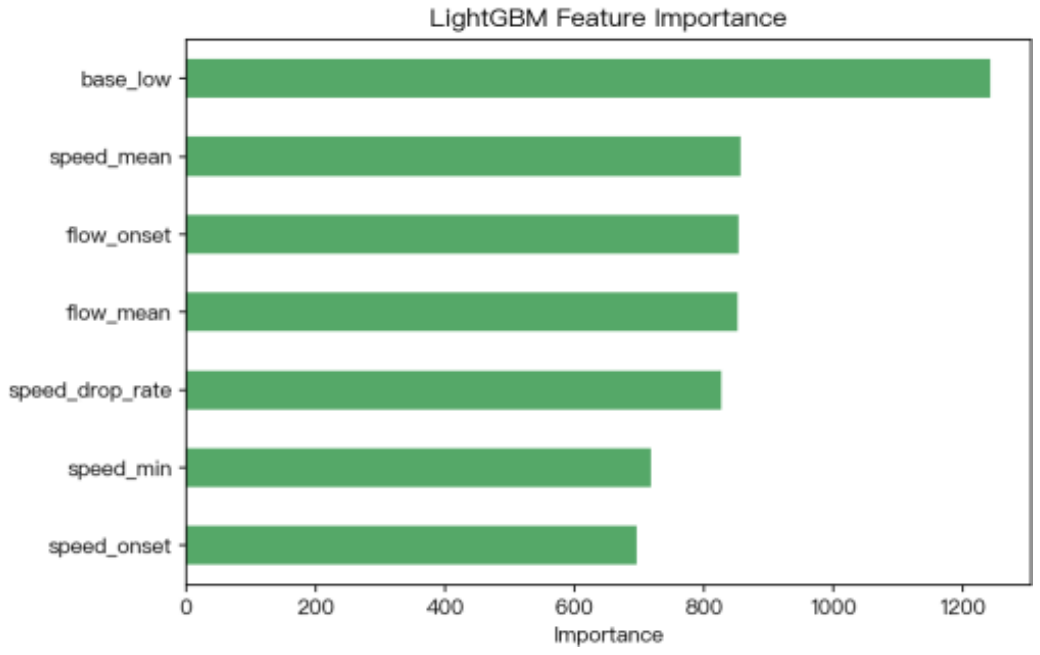

Fig. 11. Feature Importance analyse in Non-Recurrent Congestion Classification for I-210 Freeway Traffic Congestion Use Case

## V. On-Demand Distribution Evaluation Base on Machine Leaning Model

To evaluate the on-demand dissemination performance of the model in V-DTN, this study establishes a dual-simulator evaluation platform integrating SUMO and The ONE. It employs machine learning to predict congestion duration and dynamically adjusts the threshold for congestion warning dissemination in V-DTN. First, an experimental road network is constructed within SUMO. Based on the real-world I-210 Freeway Traffic Congestion Use Case—the San Gabriel Valley section of California's I-210 highway—road data are obtained from OpenStreetMap using SUMO's wizard tool. After filtering out minor roads, only the I-210 freeway and its ramps are retained, resulting in the network shown in Figure 12. Next, simulated traffic flow is introduced via the randomTrips tool. As illustrated in Figure 13, traffic congestion caused by excessive vehicle volume becomes evident. Additionally, congestion is simulated by adjusting lane speed limits and closing lanes, significantly reducing capacity at accident-prone sections and causing upstream queues that propagate backward, as shown in Figure 14. Finally, edge data with five-minute granularity is output for each lane, providing time series data on speed, flow, and occupancy, serving as synthetic counterparts to real detector data. This five-minute-granularity edge data is fed into a machine learning pipeline, which identifies 174 congestion events during the SUMO simulation. Their distribution is shown in Figure 15, where approximately one-third of all congestion events last less than 30 minutes and another third last between 30 and 60 minutes. These durations form the basis for dynamic thresholds used in on-demand V-DTN dissemination.

| **Algorithm** | **F1-Score** |
|---|---|
| Rule-based methods | 0.839 |
| LightGBM | 0.845 |

TABLE III
Performance comparison between the original model and the long-tail-removed model

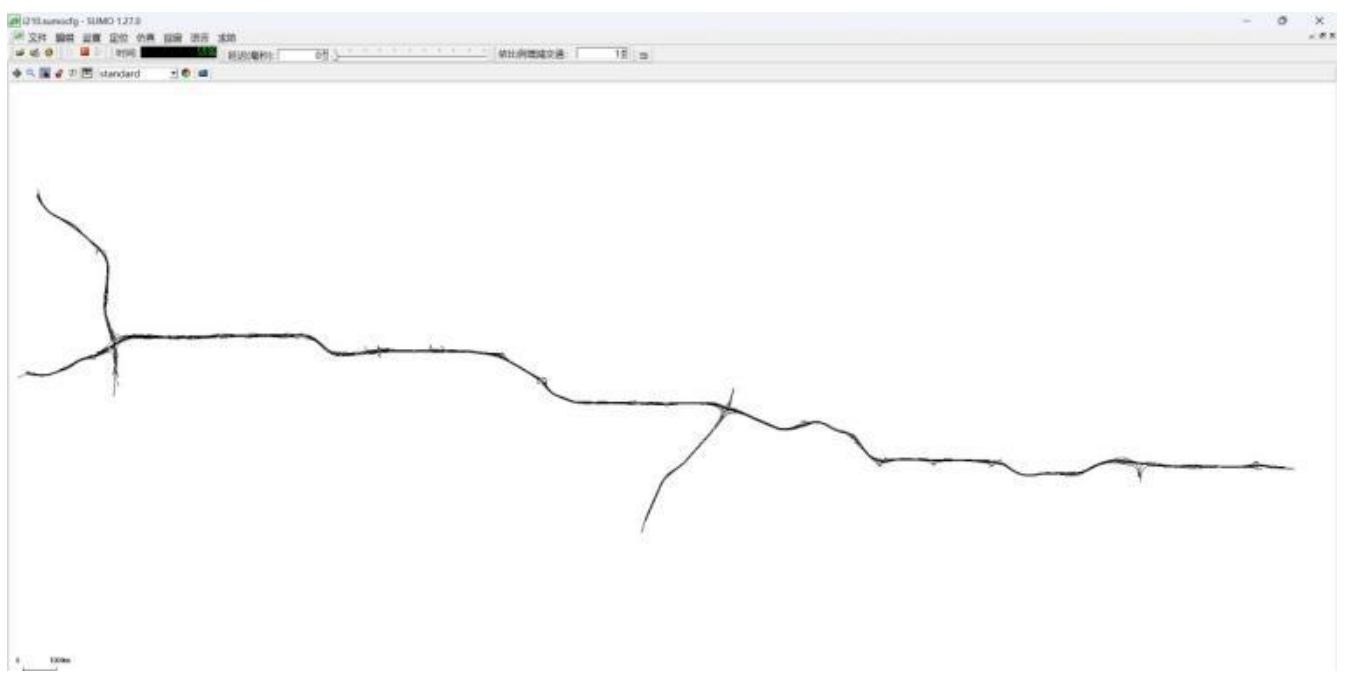

Fig. 12. I-210 Freeway Traffic Congestion Use Case in SUMO

Based on the SUMO simulation results, experiments on the efficiency of dynamic-threshold-based V-DTN dissemination are conducted in The ONE simulator. The Epidemic protocol is adopted, and due to computational overhead constraints in the network simulator, a proportional abstraction method is applied. As shown in Figure 16, six congestion points are included, each broadcasting during periods corresponding to their respective congestion durations: two lasting 20 minutes, two lasting 60 minutes, one lasting 90 minutes, and one lasting

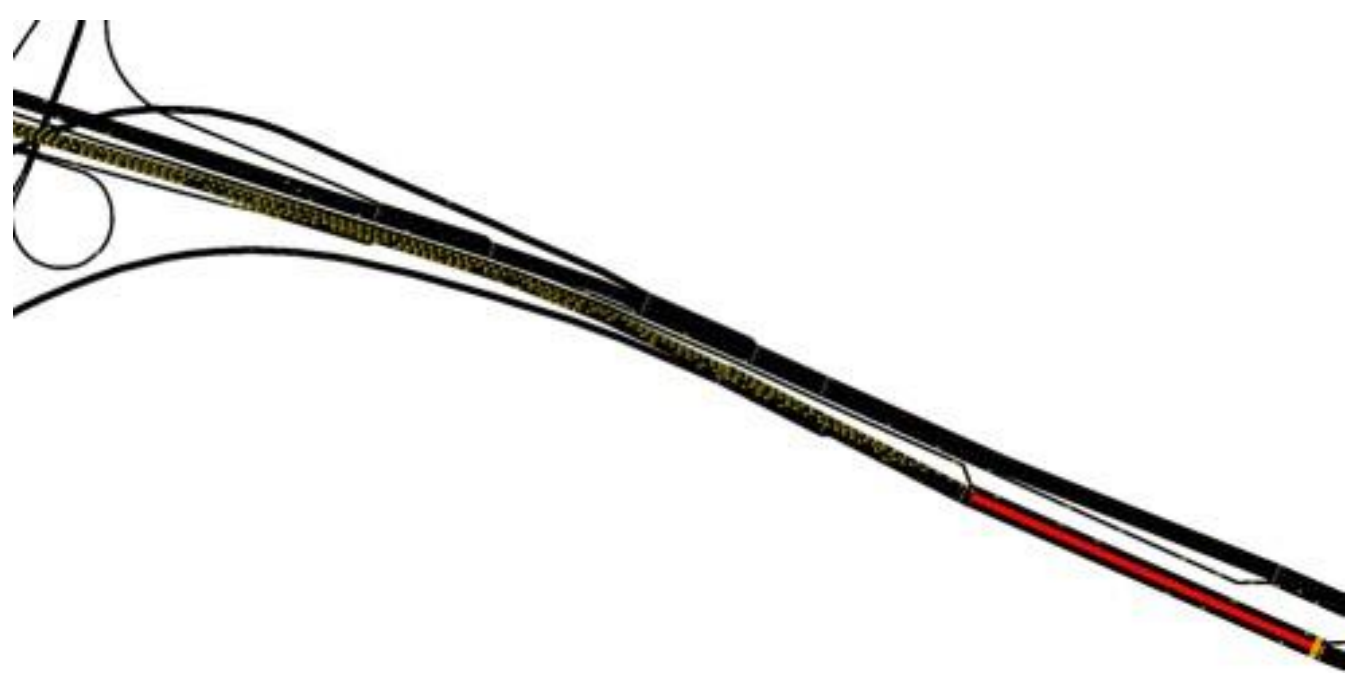

Fig. 13. SUMO Congestion Simulation in I-210 Freeway Traffic Congestion Use Case

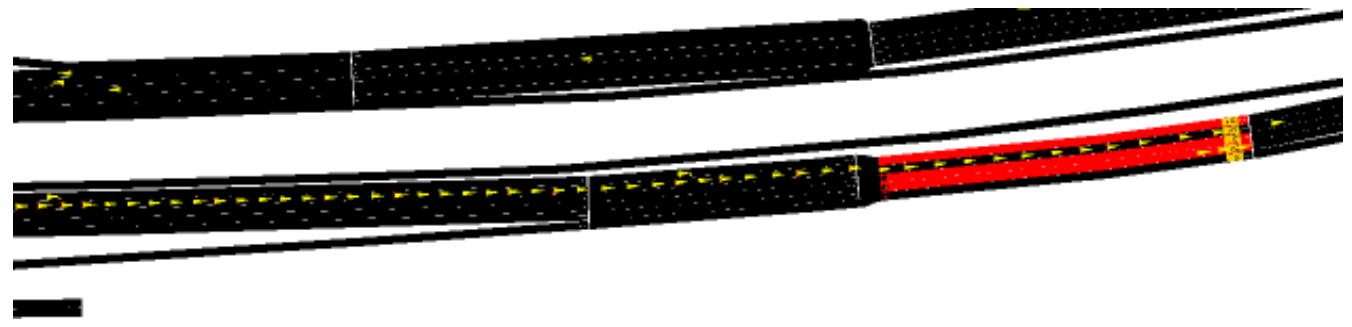

Fig. 14. SUMO Congestion Simulation in I-210 Freeway Traffic Congestion Use Case

120 minutes. Six roadside detection units (fixed nodes) are distributed across congested areas and broadcast warnings upon detecting congestion. Additionally, 90 free-flowing vehicles are simulated along ramps or opposite-direction lanes, serving both as message relays and receivers. Three simulation runs are performed: the first serves as a control group, broadcasting alerts for all congestion events; the second and third simulations apply thresholds of 30 and 60 minutes respectively, suppressing broadcasts for congestion events predicted to last shorter than the threshold.

The simulation results from The ONE demonstrate the following. First, statistics on message creation and forwarding are shown in Figure 17. As the dynamic threshold increases, message creation decreases from 8,870 to 5,043—a reduction of about 43%—while total forwarding volume drops by approximately 17%. This clearly indicates that the dynamic threshold strategy significantly reduces resource consumption in the network. Moreover, the decrease in forwarding volume is smaller than the drop in message generation. Combined with the median buffer time statistics in Figure 18, it is evident that buffer events increase with higher dynamic thresholds. This suggests that after eliminating unnecessary messages, messages remain in buffers longer, thereby improving overall forwarding efficiency. Delivery rate and delivery delay statistics further validate this conclusion, as shown in Figures 19 and 20. The delivery rate increases from approximately 0.80 to 0.86 as the dynamic threshold rises, while the median delivery delay decreases by about 38%. Simulation results show that dynamically suppressing short-term congestion broadcasts using machine learning-based congestion persistence prediction can significantly reduce network resource consumption and network layer congestion, while improving delivery reliability and reducing delivery latency.

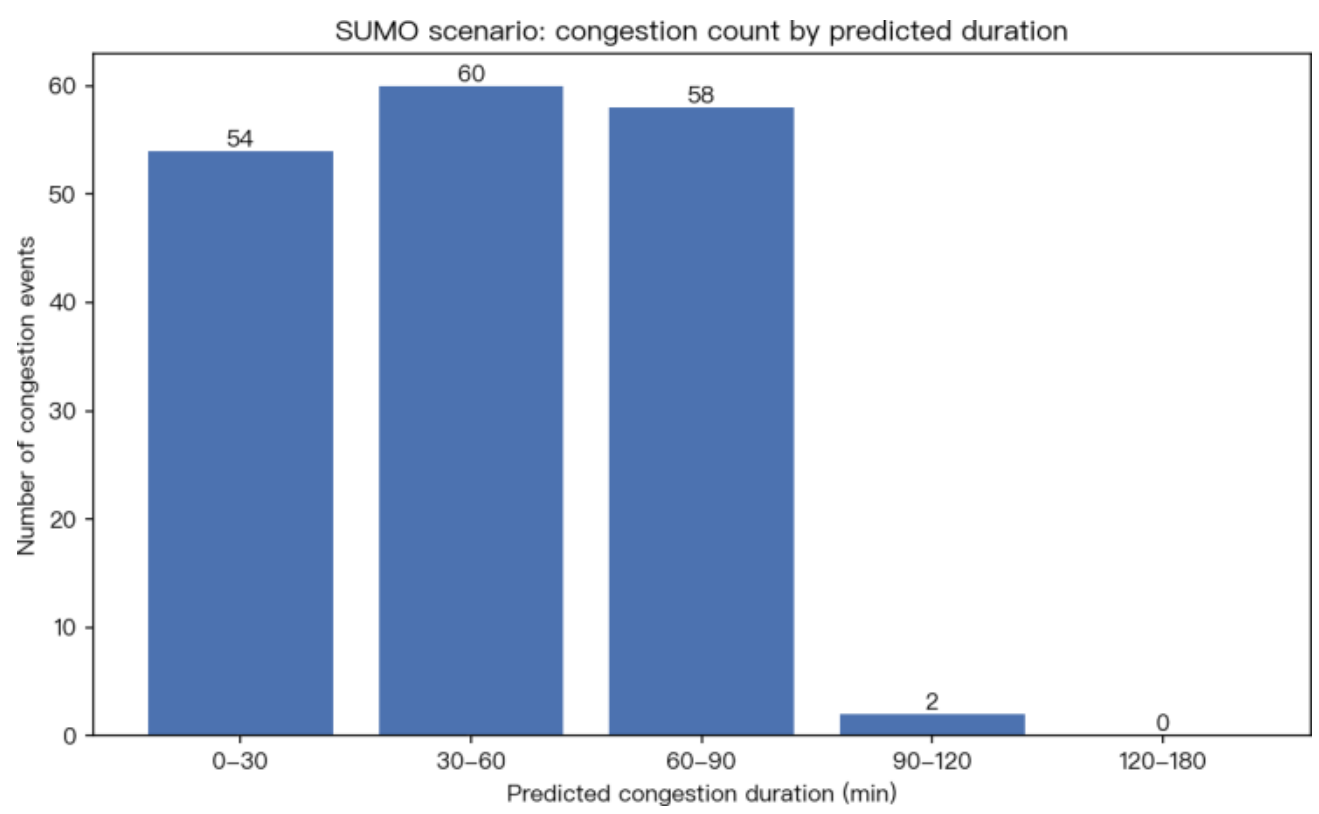

Fig. 15. Congestion Count by predicted Duration in I-210 Freeway Traffic Congestion Use Case

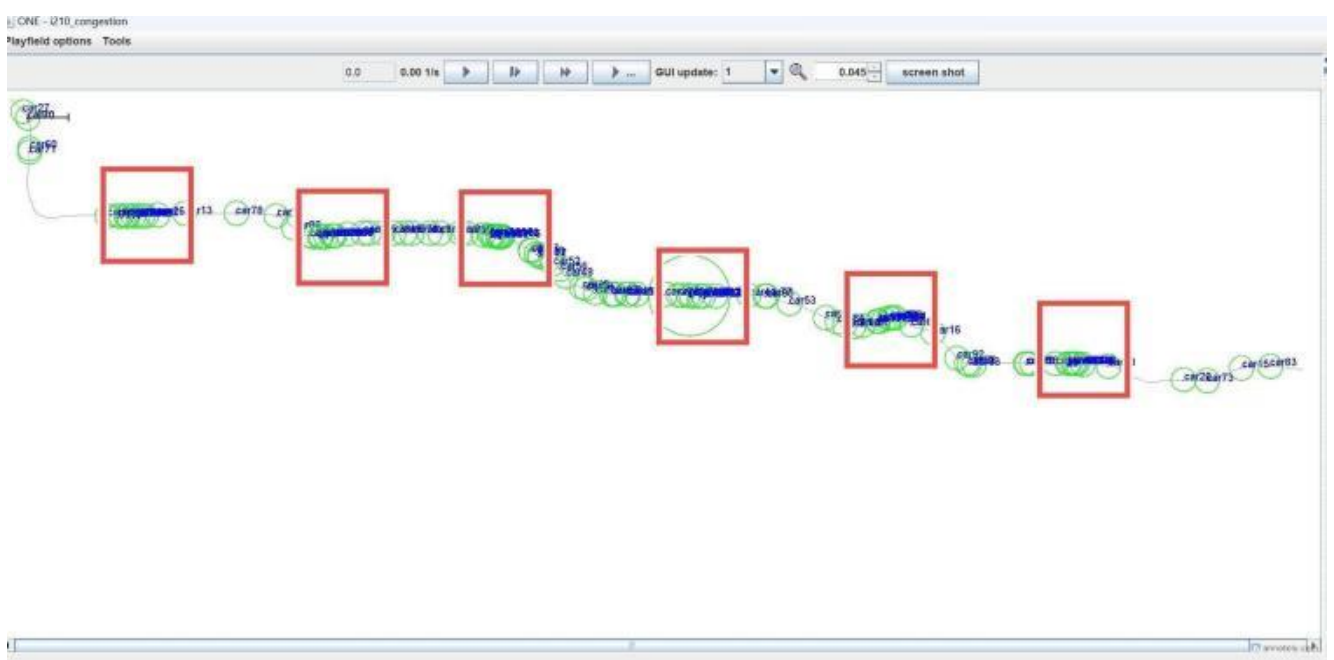

Fig. 16. Congestion area in The ONE Simulator in I-210 Freeway Traffic Congestion Use Case

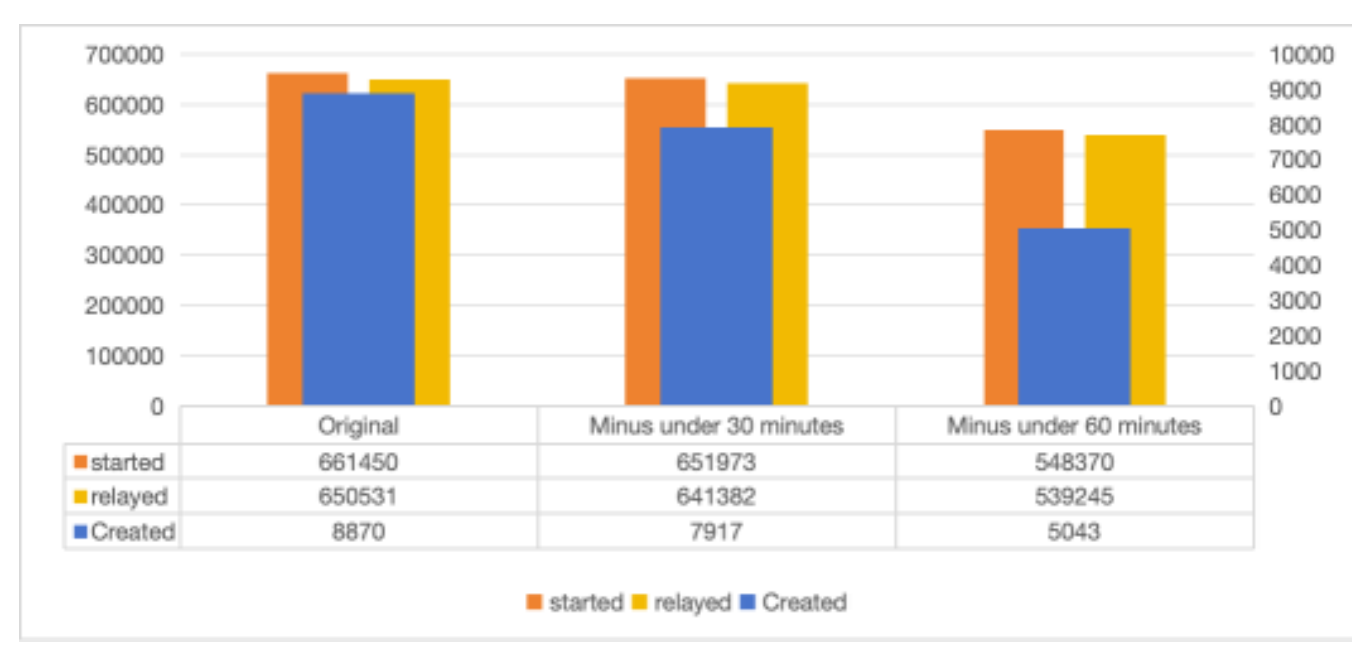

Fig. 17. Message forwarding statistics in dynamic threshold in I-210 Freeway Traffic Congestion Use Case

## VI. Conclusion and Reflections

### A. Conclusion

This project constructs the I-210 Freeway Traffic Congestion Use Case using real detector data from PeMS's I-210 corridor, proposing and validating a machine learning-based traffic congestion perception mechanism along with an event-driven dynamic threshold vehicular delay-tolerant network (DTN) on-demand message dissemination approach. In the machine learning model development, two models were trained for regression and classification tasks, respectively, targeting congestion duration and recurrence patterns. For the regression task predicting congestion duration, LightGBM achieved the best performance among compared models: it attained a 5-fold cross-validation MAE of 34.7 ± 1.1 and RMSE of 52.6 over a 240-minute sample window, and improved to MAE of 26.7 ± 0.1 and RMSE of 38.1 within an 180-minute window. Error analysis revealed that insufficient samples with congestion lasting beyond 240 minutes caused a long-tail effect, indicating that sample imbalance is a key factor leading to model underfitting when applying machine learning to congestion detection. Additionally, feature importance analysis showed that average speeds at upstream and downstream adjacent stations ranked as the top two most important features, confirming the critical role of spatial characteristics in congestion prediction. For the classification task predicting recurrent congestion, all models performed nearly identically, achieving an F1-score of approximately 0.85—showing no significant difference compared to rule-based methods without machine learning. Feature importance analysis indicated that historical free-flow speed data of lanes was the most decisive feature in determining whether congestion was non-recurrent. This result highlights that, in predicting congestion periodicity, accurate measurement and comparison of lane free-flow speeds across different day types holds greater significance than the choice of machine learning model, especially when no additional feature dimensions are introduced. Based on these machine learning results, the V-DTN simulation employed predicted congestion duration as the dynamic threshold condition for message dissemination. The experiment used SUMO to simulate congested traffic flows and applied the model for predictions, then implemented dynamic threshold suppression in message distribution via the ONE Simulator. Results showed that as the dynamic threshold increased, message generation decreased from 8,870 to 5,043—a reduction of about 43%—while total forwarding volume dropped by approximately 17%. Delivery rate improved from around 0.80 to 0.86, and median delivery delay decreased by about 38%. These findings demonstrate that integrating machine learning-based congestion perception with dynamic-threshold on-demand dissemination for short-term congestion broadcasts can significantly reduce network resource consumption and network-layer congestion, while enhancing delivery reliability and reducing delivery latency.

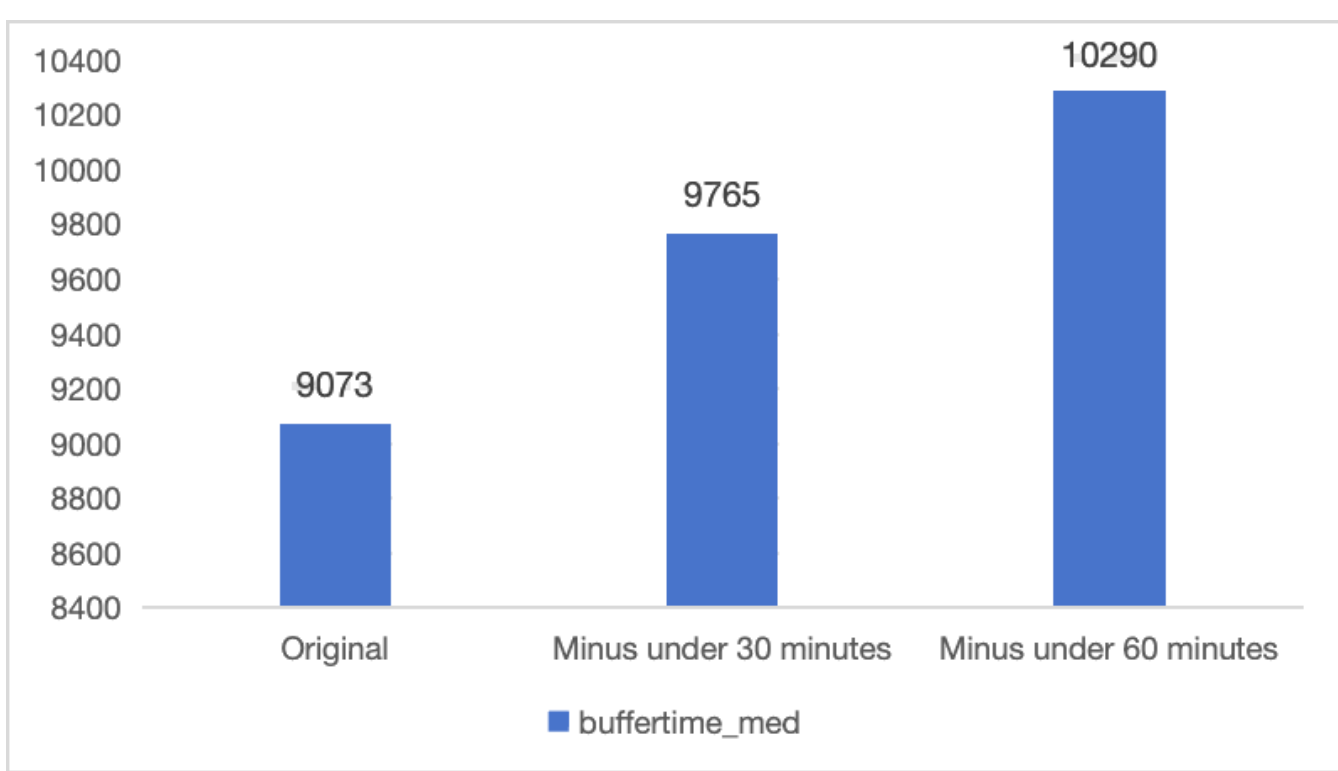

Fig. 18. Buffertime statistics in dynamic threshold in I-210 Freeway Traffic Congestion Use Case

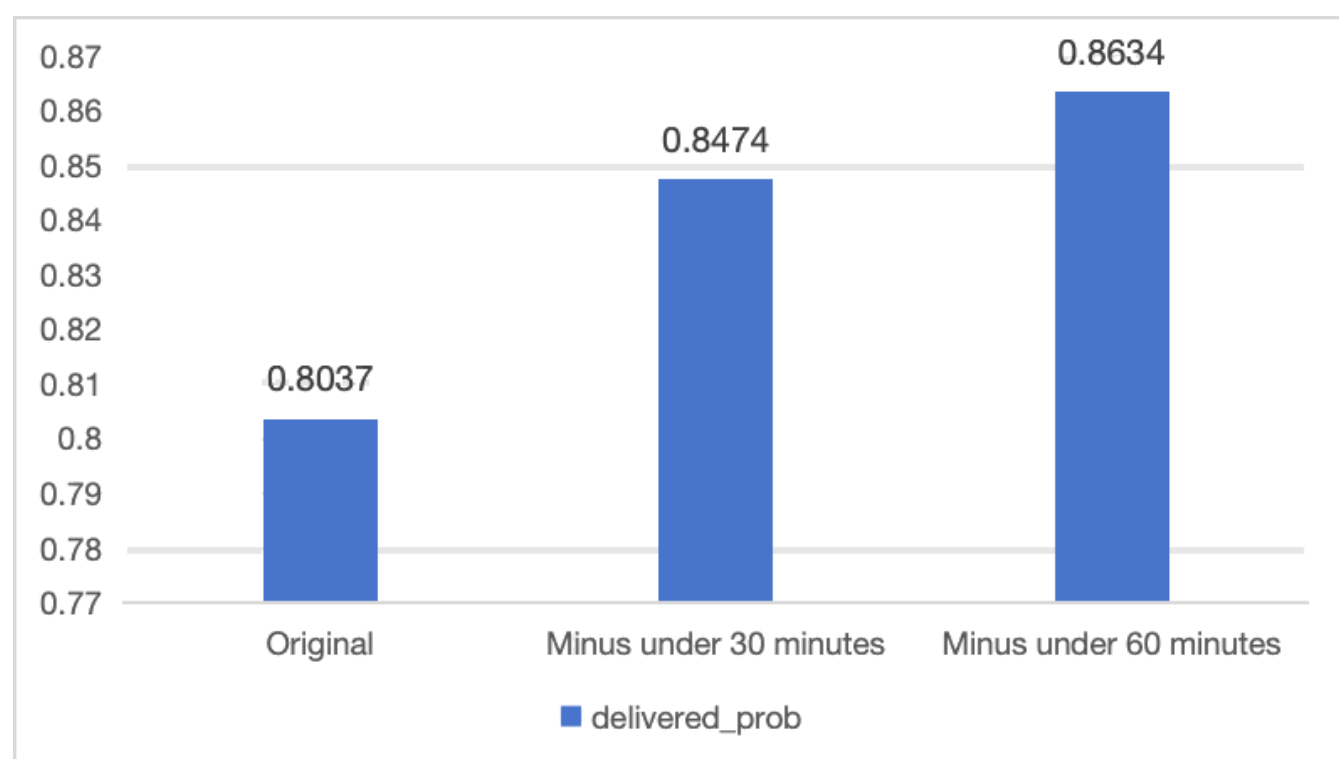

Fig. 19. Delivery probability in dynamic threshold in I-210 Freeway Traffic Congestion Use Case

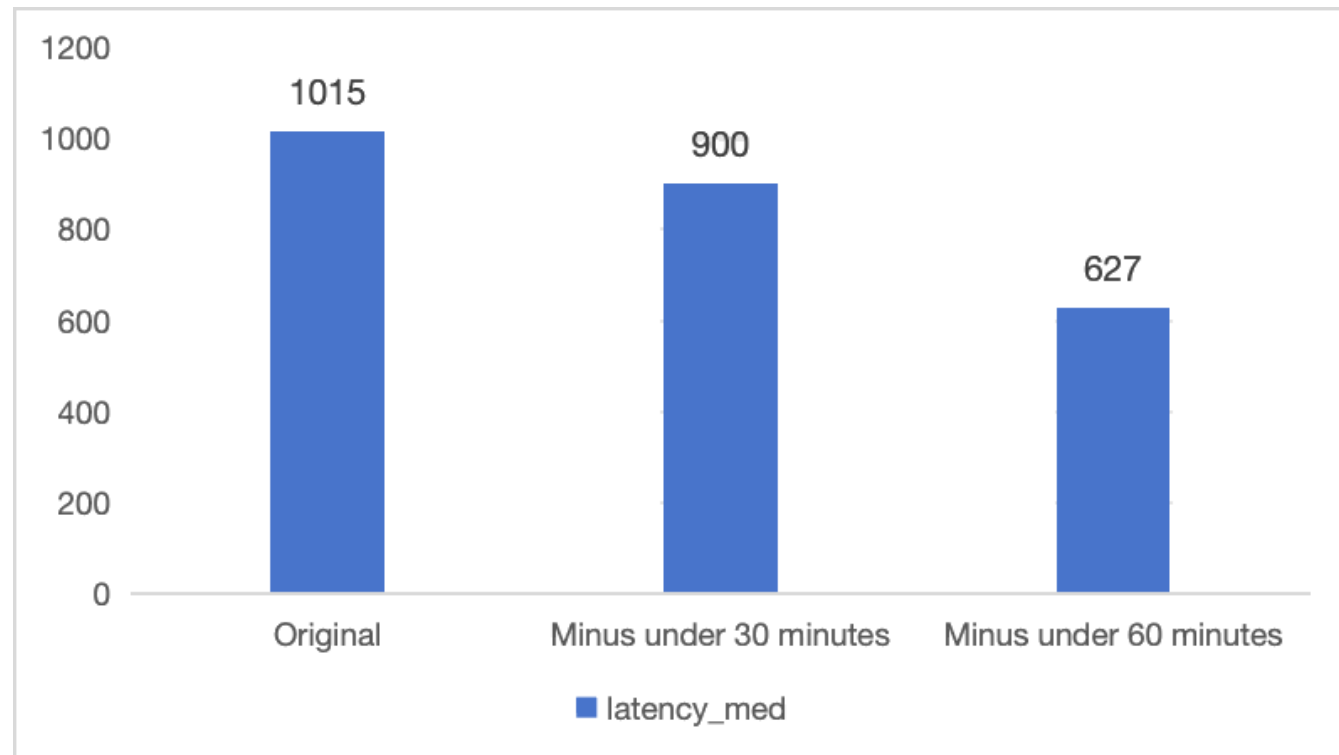

Fig. 20. Delivery latency in dynamic threshold in I-210 Freeway Traffic Congestion Use Case

### *B. Reflections and Future Work*

This project achieved a full technical chain fusing machine learning, traffic simulation, and network simulation for vehicular delay-tolerant networks (VDTNs) traffic congestion perception. It adaptively curtailed DTN dissemination based on specific congestion events, improving resource usage, congestion status, and delivery performance. Due to The ONE simulator and computing resource constraints, full SUMO vehicle trajectory transfer was incomplete, necessitating an abstracted decoupled scheme. Future work could apply the simulator integration method of [33] focusing on specific intersections to reduce hardware burden. Alternatively, utilizing

Kang et al.'s ns-3 to SUMO map model [39] or combining SUMO with OMNET++ or TrafficModeler [40] would better couple simulators and improve experiment reliability.

Algorithmically, while this project utilizes LightGBM, extending to deep learning is a key future direction. For instance, applying an RNN-RBM model can effectively forecast spatiotemporal traffic dynamics [41]. Alternatively, the Adaptive Resource Awareness and Topology Awareness Resource Provisioning Protocol (ARPP) [42] could be adopted to optimize resource provisioning in edge cloud architectures using deep reinforcement learning and LSTM. Regarding congestion prediction, the current method relies heavily on historical free-flow speeds. To improve this, upstream-downstream speed and queue length detection methods can better identify urban gridlock [43]. Furthermore, integrating traffic flow with social media data via NLP accurately detects non-recurrent congestion [44], while real-time ITS data facilitates dynamic vehicle routing [45].

Furthermore, this approach should be tested under additional DTN routing protocols with energy-saving considerations. Incorporating cognitive caching mechanisms like CafRepCache [46] and CognitiveCache [47] can improve content discovery, caching efficiency, and message propagation in dynamic opportunistic networks. For energy conservation, protocols like CognitiveCharge [48] and SmartCharge [49]—which utilize V2V/V2G energy state exchanges, historical predictions, and Q-learning—address current gaps by considering both network layer transmission and node energy states. Finally, while this study focuses on the I-210 Freeway, future implementations should explore broader urban scenarios using similar pilot corridor detector data [50]. Leveraging open-source detector data alongside spatiotemporal big data analysis, which reveals distinct seasonal and daily traffic patterns [51], will ultimately enhance the method's transferability, universality, and applicability.

## Appendix A
## Source Code

The source code for this project have been uploaded to Github. Please visit the following link. The description of the flies has been included in the README.md document. https://github.com/liuxiaofei923-tech/I-210-Freeway-Traffic-Congestion-Use-Case